\documentclass[12pt]{article}

\usepackage{amssymb}
\usepackage[dvips]{graphicx}

\newcommand\pdiv[3]{\left(\frac{\partial #1}{\partial #2}\right)_{#3}}
\newcommand\sign{\mathop{\rm sign}\nolimits}

\begin{document}

\title{Thermodynamic stability of ice models in the vicinity of a critical point}

\author{A.N.Galdina and E.D.Soldatova\\ ({\footnotesize {\it e-mail:} alexandragaldina@gmail.com, soldatovaed@gmail.com})
\\
Dnipropetrovsk National University,
Dnipropetrovsk, Ukraine}

\date{ }

\maketitle

\begin{abstract}
The properties of the two-dimensional exactly solvable Lieb and
Baxter models in the critical region are considered based on the
thermodynamic method of investigation of a one-component system
critical state. From the point of view of the thermodynamic
stability the behaviour of adiabatic and isodynamic parameters for
these models is analyzed and the types of their critical behaviour
are determined. The reasons for the violation of the scaling law
hypothesis and the universality hypothesis for the models are
clarified.\bigskip

{\it Key words:} scaling law hypotheses; universality hypotheses;
stability coefficients; critical state
\end{abstract}

\section{Introduction}

The description of the behaviour of thermodynamic parameters near
the critical points is one of the basic problems of the critical
state theory. Direct statistical calculations connected with the
evaluation of the partition function of real systems are
unavailable at present because of the impossibility of accounting
exactly for the interactions and, moreover, for the fluctuations
which are large near the critical point. So, solving the problem
by the methods of statistical physics one considers either the
simplest models, for which the partition function can be evaluated
exactly, or an approximate solution of the problem.

At the first approach the exactly solvable two-dimensional models
(the Ising, Lieb, Baxter models and others \cite{bax} forming the
most valuable possession of statistical mechanics) are of great
importance. The second approach is connected mainly with the
examination of the asymptotic behaviour of thermodynamic
parameters near the critical points, as well as with the
development of the scaling law hypothesis, the universality
hypothesis and the renormalization group approximation in various
variants and has appreciably succeeded. Indeed, the large class of
real systems and models satisfies the scaling law and the
universality hypotheses. The existence of real systems and exactly
solvable two-dimensional models, for which these hypotheses are
violated is also remarkable. The six-vertex ferroelectric Lieb
model and the eight-vertex Baxter model \cite{bax} are such
examples.

Our aim is the examination of the critical properties of these
models based on the thermodynamic method of investigation of the
critical state \cite{sold1}--\cite{sold3} which is developed on the
first principles without any hypotheses.

The method is based on the constructive critical state definition
and the critical state stability conditions. The method describes
a variety of critical state nature manifestations. The violation
of the scaling law and universality hypotheses in the Lieb and
Baxter models is explained just by this variety.

The Lieb and Baxter models give a reasonable fit to real
ferroelectrics (antiferroelectrics) and ferromagnets
(antiferromagnets). So, the application of the thermodynamic
method \cite{sold1}--\cite{sold3} to them could be interesting for
the development of the critical state theory.

\section{The thermodynamic method of investigation of the critical state}

Let us consider the basic theses of the thermodynamic method and
the terminology. The critical state definition, which considers
both the properties of homogeneous and heterogeneous system can be
written in the form \cite{sold1}--\cite{sold3}:
\begin{equation}\label{eq:a1}
\left\{\begin{array}{c}dT=\displaystyle\pdiv{T}{S}{x}dS+\pdiv{T}{x}{S}dx=0\\
dX=\displaystyle\pdiv{X}{S}{x}dS+\pdiv{X}{x}{S}dx=0\end{array}\right.,
\quad\pdiv{X}{T}{c}=-\frac{dS}{dx}=K_c.
\end{equation}
Here $X$ is the generalized thermodynamic force, $x$ is the
conjugated thermodynamic variable (the external parameter of a
system), $K_c$ is the critical slope of a phase equilibrium curve.
Eq. (\ref{eq:a1}) has non-trivial solutions, if the condition
\begin{equation}\label{eq:a2}
\left[\begin{array}{cc}\displaystyle\pdiv{T}{S}{x}&\displaystyle\pdiv{T}{x}{S}\\
\displaystyle\pdiv{T}{x}{S}&\displaystyle\pdiv{X}{x}{S}\end{array}\right]=D=
\pdiv{T}{S}{x}\pdiv{X}{x}{S}-\pdiv{T}{x}{S}^{2}=0.
\end{equation}
is fulfilled all over the spinodal. It coincides with the
well-known critical state condition $D=0$, where $D$ is the
stability determinant of the system \cite{sem1,sem2}. According to
the terminology of Refs. \cite{sem1,sem2} the parameters concerned
under the constant thermodynamic variables,
$\displaystyle\pdiv{T}{S}{x}, \pdiv{T}{x}{S}, \pdiv{X}{x}{S}$, are
the adiabatic parameters (AP's); the parameters concerned under
the constant thermodynamic forces, $\displaystyle\pdiv{T}{S}{X},
\pdiv{T}{x}{X}, \pdiv{X}{x}{T}$, are the isodynamic parameters
(IP's). The parameters $\displaystyle\pdiv{T}{S}{x}$ and
$\displaystyle\pdiv{X}{x}{S}$ are called the adiabatic stability
coefficients (ASC's); whereas $\displaystyle\pdiv{T}{S}{X}$ and
$\displaystyle\pdiv{X}{x}{T}$ are called the isodynamic stability
coefficients (ISC'c). The stability coefficients are related to
the fluctuations of the external parameters of the system (the
first and the second Gibbs lemmas) which infinitely increase near
the critical point.

The definition (\ref{eq:a1}) describes the critical state by means
of the AP's. The solution of the homogeneous linear equations
(\ref{eq:a1}) is the critical slope $K_c$ which distinguishes the
critical point on the spinodal. It is the fundamental
characteristic of the critical state and it can be expressed via
the ASC's:
\begin{equation}\label{eq:a3}
-\frac{dS}{dx}=K_c=\left[\sign\pdiv{T}{x}{S}\right]\left(
\pdiv{X}{x}{S}\pdiv{T}{S}{x}^{-1}\right)^{1/2}.
\end{equation}

This definition, being combined with the critical state stability
conditions, leads to the existence of four alternative types of
the critical behaviour of thermodynamic systems
\cite{sold1}--\cite{sold3}. The behaviour type is defined by the
value of one ASC and $K_c$.

The behaviour of the whole set of the stability characteristics of
the system (the AP's and IP's) is determined for each type. The
fourth type of the critical behaviour is the most interesting and
the most "fluctuating" one. In this case it is necessary to
consider the differential equations of higher orders. Then the
solution is realized by several possibilities
\cite{sold1}--\cite{sold3}. The case of two or even three phase
equilibrium curves converging at the critical point is of special
interest. Such a point has not yet been found experimentally, but
in this paper we demonstrate that the critical point of the
ferroelectric Lieb model has just this feature.

\section{The ferroelectric 6-vertex Lieb model}

There are a lot of crystals with the hydrogen bonds in the nature \cite{bax}.
The ions in such crystals (with the coordination number four) must
obey the ice rule. The bonds between atoms via hydrogen ions form
the electric dipoles. So, it is convenient to represent them as
the arrows on the bond curves. These arrows are directed to that
end of the bond which is occupied by the ion. There are only six
such configurations of arrows, therefore the ice models are
sometimes called the six-vertex models. The partition function of
such a system is defined by the expression
\begin{equation}\label{eq:a4}
Z=\exp
[-(n_1\varepsilon_1+n_2\varepsilon_2+\ldots+n_6\varepsilon_6)/kT],
\end{equation}
where the summation should be carried out over all the
configurations of the hydrogen ions allowed by the ice rule,
$\varepsilon_i$ is the energy of $i$-type vertex configuration and
$n_i$ is the number of $i$-type vertices in the lattice.

There are three sorts of the ice models which have been solved by
E. H. Lieb \cite{lieb1,lieb2}. One of them, considered in this
paper, can describe $KH_2PO_4$ (KDP), the crystal with hydrogen
bonds, which is characterized by the coordination number four and
orders ferroelectrically at low temperatures under the appropriate
choice of $\varepsilon_1, \varepsilon_2, \ldots, \varepsilon_6$.
For the square lattice this choice is
\begin{equation}\label{eq:a5}
\varepsilon_1=\varepsilon_2=0, \
\varepsilon_3=\varepsilon_4=\varepsilon_5=\varepsilon_6>0.
\end{equation}
In the ground state all the arrows are directed either up and to
the right or down and to the left. Both these states are typical
for the ordered ferroelectric.

\begin{figure}
\begin{center}
\includegraphics[width=10cm]{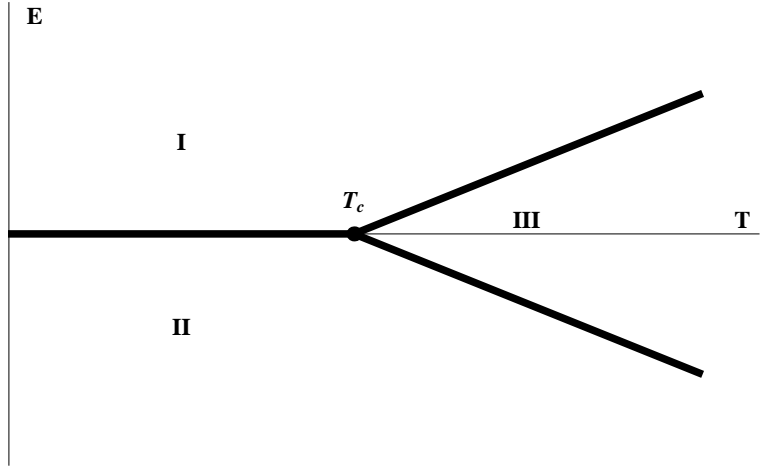}
\end{center}
\caption{The phase diagram of Lieb model \cite{bax}}\label{phase}
\end{figure}

The expression for the free energy per lattice point in the
presence of the nonzero external field is given by
\begin{equation}\label{eq:a6}
f=\varepsilon_1-EP-\frac 1 2 k(T-T_c)(1-P^2)+
A\left[\frac{T-T_c}{T_c}\right]^{3/2},
\end{equation}
where $P$ is the electric polarization \cite{bax}, and $A=-0.2122064 k T_c$; $k$ is Boltzmann constant. The critical
equation of state is expressed in the form
\begin{equation}\label{eq:a7}
P=\left\{\begin{array}{cc}\displaystyle\frac{E}{k(T-T_c)},&
\mbox{if}\
|E|<k(T-T_c)\\
\sign (E) & \mbox{otherwise}.\end{array}\right.
\end{equation}
It corresponds to the phase diagram in Fig. \ref{phase}.

It is necessary to emphasize that the ice model allows the
investigation on the basis of the thermodynamic method. In this
case the temperature $T$ and the electric intensity $E$ stand for
the generalized thermodynamic forces. The conjugated generalized
thermodynamic variables are the entropy $S$ and the electric
polarization $P$. Thus, the adiabatic parameters for the given
model are $\displaystyle\pdiv{T}{S}{P}, \pdiv{T}{P}{S}$ and
$\displaystyle\pdiv{E}{P}{S}$, and the isodynamic parameters are
$\displaystyle\pdiv{T}{S}{E}, \pdiv{T}{P}{E}$ and
$\displaystyle\pdiv{E}{P}{T}$. As $T\rightarrow T^+_c$ the free
energy per lattice point coincides with expression (\ref{eq:a6}),
and as $T\rightarrow T^-_c$ the free energy equals simply to
$\varepsilon_1-EP$. Consequently, the heat capacity is finite in
the subcritical region and the critical exponent is $\alpha'=0$.
Both the phases are quite ordered and then differ from each other
only by a direction of the electric polarization vector ($P=\pm
1$). This corresponds to the second critical behaviour type
according to the thermodynamic classification of critical
behaviour types of one-component systems \cite{sold2}:
$\displaystyle\pdiv{T}{S}{P}=\frac{T}{C_{P}}\neq\{0, \infty\},\
\pdiv{E}{P}{S}=0$. Thus, the critical slope of the equilibrium
curve of the phases $I$ and $II$ equals to zero, $K_c=0$.

As we can see from Eq. (\ref{eq:a6}), in the supercritical region
($T\rightarrow T_c^+$) the heat capacity diverges as
$\displaystyle\left(\frac{T-T_c}{T_c}\right)^{-1/2}$, i.e. the
thermic ASC is
$\displaystyle\pdiv{T}{S}{P}=C\sqrt{\frac{T-T_c}{T_c}}$. Let us
approach to the critical point from the supercritical region along
the curve of the first-kind phase transition $I-III$ and $II-III$.
It is known that at least one of the jumps $\Delta P$, $\Delta S$
must exist along these curves. I.e., on the transition curve
\begin{equation}\label{eq:a8}
\Delta P=P_I-P_{III}=1-\frac{E}{k(T-T_c)}\neq 0.
\end{equation}
At the critical point $\Delta P=0$.

The entropy jump can be determined from the known behaviour of the
heat capacity. For the phase $I$ we have $\alpha'=0$, i.e.
$C_P=const$. Consequently, the entropy of the phase $I$ is
$S_I=C_1\ln T+const$. For the phase $III$ we have $\alpha=1/2$,
i.e. $S_{III}=C_2\sqrt{T_c(T-T_c)}+const$. Then, for the jump, we
have
\begin{equation}\label{eq:a9}
\Delta S=S_I-S_{III}=C_1\ln T-C_2\sqrt{T_c(T-T_c)}+const.
\end{equation}
At the critical point $\Delta S=const\neq\{0, \infty\}$. Such a
behaviour of the entropy is connected with the divergence of the
heat capacity in the supercritical region.

The analogous results can be obtained for phases $II-III$ as well.

For the equilibrium line $I-II$ we have $\Delta P=2$, $\Delta
S=0$. At the critical point $\Delta P=0$.

Thus, the found values of the jumps correspond to the results of
papers \cite{sold1}--\cite{sold3}, and the point $T_c$ is critical
for the phase equilibrium line $I-II$, for the line $I-III$ and
for the line $II-III$. So, the point $C$ in the phase diagram
(Fig. \ref{phase}) is the point of the convergence of three phase
equilibrium lines. The possibility of such a point has been
predicted in papers \cite{sold1}--\cite{sold3}.

Let us analyze the behaviour of the whole set of the system
stability characteristics (the AP's and the IP's). The relations
between the adiabatic and isodynamic parameters exist:
\begin{equation}\label{eq:a10}
\pdiv{T}{S}{P}\pdiv{E}{P}{T}=\pdiv{T}{S}{E}\pdiv{E}{P}{S}=-\pdiv{T}{P}{S}\pdiv{T}{P}{E}.
\end{equation}
Using Eqs. (\ref{eq:a7}) and (\ref{eq:a10}), we can obtain the
following expressions for the AP's and the IP's:
\begin{equation}\label{eq:a11}
\begin{array}{cc}
\displaystyle\pdiv{T}{S}{P}=C\sqrt{\frac{T-T_c}{T_c}}, &
\displaystyle\pdiv{E}{P}{S}=\frac{k\sqrt{T_c(T-T_c)^3}}{kP^2+C\sqrt{T_c(T-T_c)}},\\
\displaystyle\pdiv{T}{P}{S}=\frac{kP(T-T_c)}{kP^2+C\sqrt{T_c(T-T_c)}},
&
\displaystyle\pdiv{T}{S}{E}=\frac{T-T_c}{kP^2+C\sqrt{T_c(T-T_c)}},\\
\displaystyle\pdiv{E}{P}{T}=k(T-T_c), &
\displaystyle\pdiv{T}{P}{E}=-\frac{T-T_c}{P},
\end{array}
\end{equation}
where $C=-\frac{4T^2_c}{3A}=6.2831908\cdot\frac{T_c}{k}$.
The critical slope equals $K^{(1)}_c=kP$ for the line $I-III$ and
$K^{(2)}_c=-kP$ for the line $II-III$.

As it follows from Eq. (\ref{eq:a11}), at $T\rightarrow T^+_c$ all
the thermodynamic stability characteristics tend to zero:
\[\begin{array}{ccc}
\displaystyle\pdiv{T}{S}{P}\rightarrow 0, &
\displaystyle\pdiv{E}{P}{S}\rightarrow 0, &
\displaystyle\pdiv{T}{P}{S}\rightarrow 0, \\
\displaystyle\pdiv{T}{S}{E}\rightarrow 0, &
\displaystyle\pdiv{E}{P}{T}\rightarrow 0, &
\displaystyle\pdiv{T}{P}{E}\rightarrow 0.
\end{array}\]

According to the critical behaviour classification \cite{sold2} at
$K_c\neq \{0, \infty\}$ and ASC's$\rightarrow 0$ we have the
fourth type of the critical behaviour, and two phase equilibrium
lines with different critical slopes $K_c^{(1,2)}=\pm kP$ converge
at the critical point. This behaviour type is the most fluctuating
one (the fluctuations of the energy and polarization
$\overline{(\Delta H)^2}, \overline{(\Delta P)^2}\rightarrow
\infty$). Approaching to the critical point from the subcritical
region (along the phase equilibrium line $I-II$ with the slope
$K_c=0$) the second type of critical behaviour is realized (the
fluctuations of the energy $\overline{(\Delta H)^2}$ is finite and
the fluctuations of the polarization $\overline{(\Delta
P)^2}\rightarrow\infty$).

As it is known, stability characteristics are inversely proportional to fluctuations
of external parameters of the system. At the continuous
transitions \cite{sem2} $D$ and the SC's pass finite minima, that corresponds
to the growth of fluctuations. The locus of these minima is curve
of supercritical transitions (the lowered stability curve or
quasispinodal). The limit case of these continuous transitions,
when fluctuations in the system are at the high and $D$ and the
SC's pass zero minima, is the critical state. The critical point
is also the limit point of some first-kind transition (the limit
point of phase equilibrium curve). If the phase equilibrium curve and curve
of supercritical transitions pass into each other continuously, i.e. the slopes of these curves are the same, then the tricritical point is observed, where three phases become identical: two subvritical phases and supercritical one.

On the quasispinodal the next condition is fulfilled \cite{cmp}:
\begin{equation}\label{eq:qq}
dD=\pdiv{D}{S}{x}dS+\pdiv{D}{x}{S}dx=0,\end{equation}
or, equivalently,

\[
\pdiv{D}{S}{x}=0,\quad\pdiv{D}{x}{S}=0.
\]

\begin{figure}
\begin{center}
\includegraphics[width=10cm]{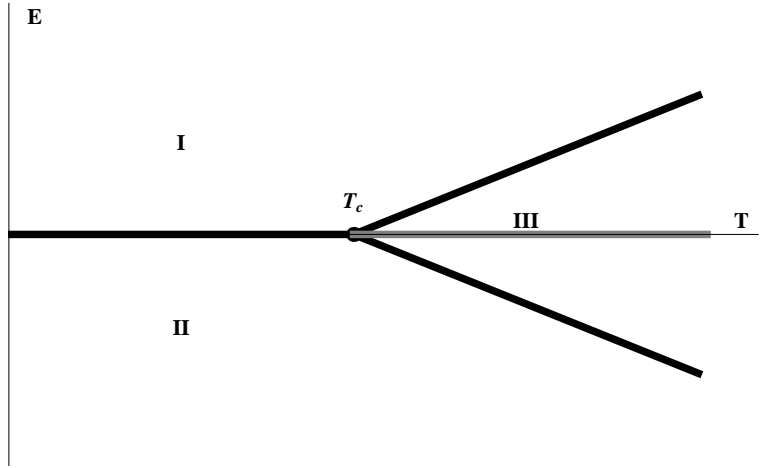}
\end{center}
\caption{The quasispinodal for Lieb model (the gray line)}\label{phase1}
\end{figure}

Using results (\ref{eq:a11}) to find the determinant of stability for Lieb model, and investigating where condition (\ref{eq:qq}) is fulfilled, we obtain $E=0$. This is equation of quasispinodal for ferroelectric Lieb model. The resulting phase diagram is shown in Fig. \ref{phase1}. So, the maximal growth of fluctuations is observed under
zero electric field. The critical slope of the subcritical phase
equilibrium curve is $K_c=0$. It means, that for this model it is
realized the case of continuous passage of the equilibrium curve
into the lowered stability curve because of the same critical
slopes.

Thus, the violation of the scaling law hypothesis in the Lieb
model can be explained by the fact that the model corresponds to
two different critical behaviour types: at $T\rightarrow T_c^+$
the second type and at $T_c^-$ the fourth type is fulfilled.
Besides, the critical point of the Lieb model is the critical
point of a special type with the convergence of three phase
equilibrium lines. Moreover, the equilibrium curve continuously passes
into the lowered stability curve.

\section{8-vertex Baxter model}

The eight-vertex Baxter model is a generalization of the
six-vertex Lieb model. The ice models as the models of the
critical phenomena have some unusual properties: the ferroelectric
state at these models is frozen (i.e. there is complete ordering
even at the non-zero temperature); the critical behaviour of
antiferroelectrics is characterized by the more complicated law
instead of a simple power dependence of $(T-T_c)$.

The first of these unusual properties is connected with the ice
structure. In the case of an unlimited lattice with the
ferroelectric ordering the infinite energy is needed for a
deformation. So, the deformation gives an infinitesimal
contribution to the partition function \cite{bax}.

The following generalization of the ice-type models was proposed
\cite{bax1}--\cite{fan}:
\begin{itemize}
\item there is only one arrow on each square lattice edge;

\item the configurations with an even number
   of arrows getting in (and getting off) each vertex are allowed only;

\item eight possible configurations of the arrows to a vertex exist.
\end{itemize}

The formation of $j$-type vertex needs the energy $\varepsilon_j$
(where $j=1,...,8$).

For such a model the partition function is given by (\ref{eq:a4})
where the summation is performed over the eight vertex
configurations.

Thus, besides the first six vertices coinciding with the Lieb
model there are another two new vertices for which all the arrows
get either in a vertex or off a vertex. Now the finite energy is
needed for the local deformation of the lattice state (e.g. for
reversing of all the arrows which are lying on the square side),
in which all the arrows are directed up or to the right. So the
ferroelectric state is ordered not completely.

As it was mentioned above, the Baxter model is fitted to describe
the critical phenomena in ferroelectrics (antiferroelectrics). The
eight-vertex model can be considered also as two Ising models with
the nearest neighbours interaction (each model is on its
sublattice). These sublattices are connected by means of the
four-spin interplay. In this case the model corresponds to
ferromagnets.

The Baxter model has the exact solution only in the absence of an
external field. The critical exponents of this model equal \cite{bax}
\begin{equation}\label{eq:a12}
\begin{array}{ccc}
\alpha=\alpha'=2-\displaystyle\frac{\pi}{\mu}, & & \\
\displaystyle\beta=\frac{\pi}{16\mu}, &
\displaystyle\gamma=\frac{7\pi}{8\mu}, &
\delta=15,\\
\displaystyle\beta_{e}=\frac{\pi-\mu}{4\mu}, &
\displaystyle\gamma_{e}=\frac{\pi+\mu}{2\mu}, &
\displaystyle\delta_{e}=\frac{3\pi+\mu}{\pi-\mu}.
\end{array}
\end{equation}
Here the index $e$ denotes the electric exponents. The exponents
$\beta$, $\gamma$ and $\delta$ are related to ferromagnet. The
exponent $\alpha$ is the same both for the ferromagnet and for the
ferroelectric. $\mu$ is the interaction parameter, it takes a
value from $(0, \pi)$. Thus, as we can see, the critical exponents
depend on the interaction parameter continuously. This fact is in
the contrary to the universality hypothesis. This result
distinguishes the Baxter model among the others two-dimensional
exactly solvable models. Taking this into the consideration, one
should expect that the type of the critical behaviour according to
the thermodynamic classification and the value of the critical
slope change depending on the interaction parameter. Let us show
this.

\subsection{The ferromagnetic Baxter model}

In the case of ferromagnet the adiabatic stability coefficients
get the following asymptotic form:
\[\pdiv{T}{S}{M}\sim t^{\displaystyle 2-\frac{\pi}{\mu}},
\quad\pdiv{H}{M}{S}\sim t^{\displaystyle\frac{7\pi}{8\mu}},\]
where $\displaystyle t=\frac{T-T_{c}}{T_{c}}$. It is necessary to
note that in the absence of the external field the behaviour of
the isodynamic parameters coincides with the behaviour of the
adiabatic parameters. When
$0<\mu\leqslant\displaystyle\frac{\pi}{2}$ the exponent $\alpha$
is negative, the exponent $\gamma$ takes a positive value, i.e.
\[\pdiv{T}{S}{M}\neq 0,\ \pdiv{H}{M}{S}=0\ \Rightarrow\ \pdiv{T}{M}{S},\
K_c=0\] and the second type of critical behaviour is fulfilled. At
$\displaystyle\frac{\pi}{2}<\mu<\frac{15\pi}{16}$  the fourth type
of critical behaviour is realized, the exponent $\alpha$ increases
($\displaystyle 0<\alpha<\frac{14}{15}$) and the exponent $\gamma$
decreases ($\displaystyle\frac{7}{4}>\gamma>\frac{14}{15}$), where
$\alpha<\gamma$.

\begin{figure}[h]
\begin{center}
\includegraphics[width=120mm]{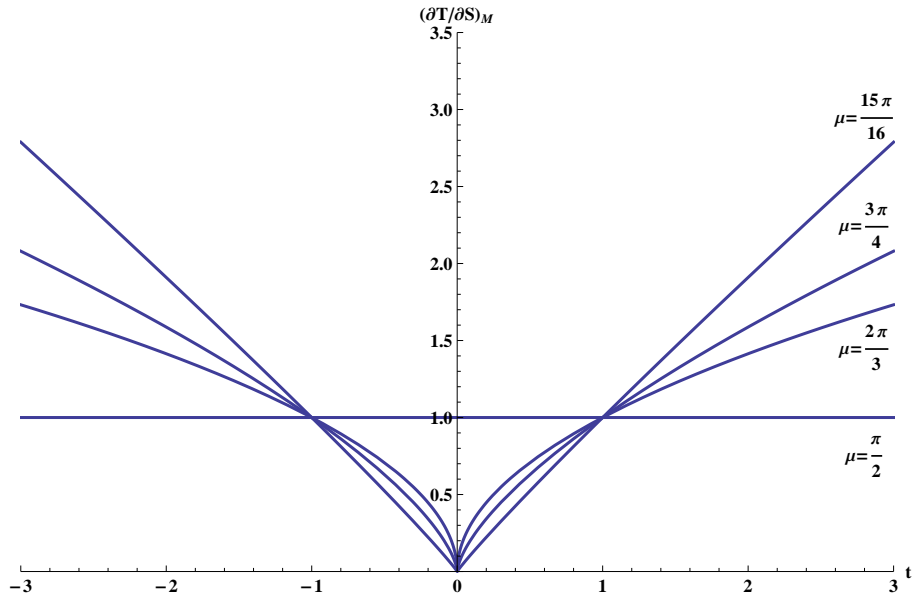}
\caption{The temperature dependence for reduced thermic coefficient of stability}\label{baxts}
\end{center}
\end{figure}
\begin{figure}[h]
\begin{center}
\includegraphics[width=120mm]{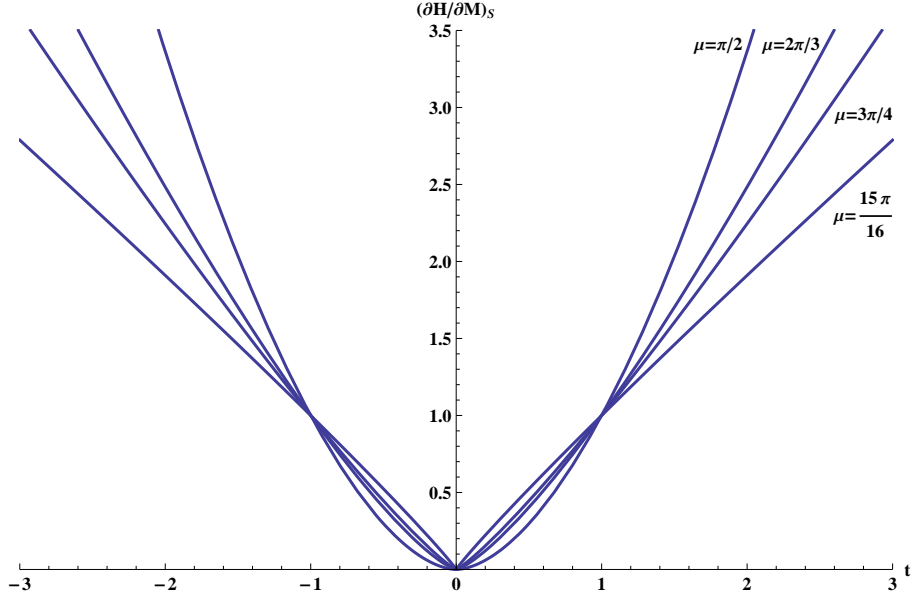}
\caption{The temperature dependence for reduced magnetic coefficient of stability}\label{baxhm}
\end{center}
\end{figure}

$$\pdiv{T}{S}{M}=0, \ \pdiv{H}{M}{S}=0 \Rightarrow \pdiv{T}{M}{S}=0.$$
All the parameters in this case tend to zero, but
$\displaystyle\pdiv{H}{M}{S}$ and $\displaystyle\pdiv{H}{M}{T}$
tend to zero faster than other parameters. The value of the
critical slope is $K_c=0$. The case
$\mu=\displaystyle\frac{15\pi}{16}$ corresponds also to the fourth
type of critical behaviour, but
$\alpha=\gamma=\displaystyle\frac{14}{15}$ and all the parameters
tend to zero according to the same law, the critical slope is
$K_c\neq\{0, \infty\}$. At $\displaystyle\frac{15\pi}{16}<\mu<\pi$
the fourth type of critical behaviour is also fulfilled,
$\displaystyle\frac{14}{15}<\alpha<1$ and
$\displaystyle\frac{14}{15}>\gamma>\frac{7}{8}$ and everywhere
$\alpha>\gamma$. All the parameters tend to zero, but
$\displaystyle\pdiv{T}{S}{M}$ and $\displaystyle\pdiv{T}{S}{H}$
tend to zero faster than other parameters. The value of the
critical slope is $K_c=\infty$. The corresponding plots of ASC's for various $\mu$ are presented in Figs. \ref{baxts}, \ref{baxhm}.

Thus, the performed analysis has determined that at $\displaystyle
0<\mu\leqslant \frac{\pi}{2}$ the critical behaviour of the Baxter
model corresponds to the second type according to the
thermodynamic classification \cite{sold1}--\cite{sold3} with
$K_c=0$, and at $\displaystyle\frac{\pi}{2}<\mu<\pi$ it
corresponds to the fourth type which is fulfilled by three
possibilities for the critical slope ($K_c=0,\ K_c\neq \{0,
\infty\}, K_c=\infty$) depending on the value of $\mu$ varying
within the mentioned interval.

\subsection{The ferroelectric Baxter model}

In the case of the ferroelectric Baxter model the stability
coefficients can be written in the form:
$$\pdiv{T}{S}{P}\sim t^{\displaystyle 2-\frac{\pi}{\mu}},
\quad \pdiv{E}{P}{S}\sim t^{\displaystyle\frac{\pi+\mu}{2\mu}}.$$
At $\displaystyle 0<\mu\leqslant\frac{\pi}{2}$, as in the previous
case, $\alpha$ is negative and $\gamma$ is positive. So
$\displaystyle\pdiv{T}{S}{P}\neq 0,\ \pdiv{E}{P}{S}=0\
\Rightarrow\ \pdiv{T}{P}{S}=0$, $K_c=0$ and the second type of
critical behaviour is fulfilled. At
$\displaystyle\frac{\pi}{2}<\mu<\pi$ the exponent $\alpha$ takes
positive values $0<\alpha<1$, but $\alpha$ is less than $\gamma$,
$\displaystyle\frac{3}{2}>\gamma>1$ and the fourth type of
critical behaviour with $K_c=0$ is realized.

It is interesting to emphasize the fact that for real ferromagnets
and ferroelectrics the critical behaviour types are also the
second and the fourth ones.

\section{Conclusion}

Thus, in the paper the consideration of the thermodynamic
stability of the Lieb and Baxter models by the method of Ref.
\cite{sold1}--\cite{sold3} has been performed. The asymptotic
expressions for the whole set of the stability characteristics are
determined.The reasons for the violation of the scaling law and
universality hypotheses in the given models are clarified. So, we
determine that the second and the fourth type of critical
behaviour is fulfilled in the subcritical and in the supercritical
region of the Lieb model, correspondingly. The violation of the
scaling law hypothesis in the ferroelectric Lieb model can be
explained just by the difference of the behaviour types. It has
been also ascertained that three phase equilibrium lines with
different critical slopes converge at the critical point of the
model. A possibility of the existence of such a type of the
critical point has been predicted in papers
\cite{sold1}--\cite{sold3}. The equation of quasispinodal is obtained and it is shown that  the equilibrium curve continuously passes
into the lowered stability curve in this model.

In the Baxter model the fulfillment of
the second and the fourth type of critical behaviour also occurs,
moreover, the fourth type is represented by three possibilities
--- with three different critical slopes of the phase equilibrium
line. The reason for the violation of the universality hypothesis
is that each of the mentioned types (the second type, the fourth
type with $K_c=0$, the fourth type with $K_c\neq\{0, \infty\}$ and
the fourth type with $K_c=\infty$) is connected either to the
certain value or the continuous range of the interaction parameter
$\mu$. It is interesting to emphasize that in each model while one
hypothesis is violated the another one is nevertheless holds. In
addition, the special case of the eight-vertex Baxter model, in
which the universality hypothesis is violated, is the Lieb model
($\mu=0$), where the universality hypothesis is satisfied, but the
scaling law hypothesis is violated, and the Ising model
($\displaystyle\mu=\frac{\pi}{2}$), where both hypotheses are
fulfilled. Therefore, the abilities of the thermodynamic method of
investigation of the one-component system critical state have been
illustrated by the example of the above-mentioned models and the
global reasons for the violation of the scaling law and
universality hypotheses concerned with the variety of the critical
state nature manifestation are revealed.

%
%


\end{document}